# Logarithmic Elastic Response in the Dilute non-Kramers System Y$_{1-x}$Pr$_x$Ir$_2$Zn$_{20}$

Tatsuya Yanagisawa[1*], Hiroyuki Hidaka[1], Hiroshi Amitsuka[1], Sergei Zherlitsyn[2], Joachim Wosnitza[2,3], Yu Yamane[4], and Takahiro Onimaru[4]

[1]*Department of Physics, Hokkaido University, Sapporo 060-0810, Japan*
[2] *Hochfeld-Magnetlabor Dresden (HLD-EMFL) and Würzburg-Dresden Cluster of Excellence ct.qmat, Helmholtz-Zentrum Dresden-Rossendorf, 01328 Dresden, Germany*
[3] *Institut für Festkörper- und Materialphysik, TU Dresden, 01062 Dresden, Germany*
[4] *Graduate School of Advanced Sciences of Matter, Hiroshima University, Higashi-Hiroshima 739-8530, Japan*

*E-mail: tatsuya@phys.sci.hokudai.ac.jp

Ultrasonic investigations of the single-site quadrupolar Kondo effect in diluted Pr system Y$_{0.966}$Pr$_{0.034}$Ir$_2$Zn$_{20}$ are reported. The elastic constant $(C_{11}-C_{12})/2$ is measured down to ~40 mK using ultrasound for the dilute system Y$_{0.966}$Pr$_{0.034}$Ir$_2$Zn$_{20}$ and the pure compound YIr$_2$Zn$_{20}$. We found that the elastic constant $(C_{11}-C_{12})/2$ of the Pr-dilute system exhibits a logarithmic temperature dependence below $T_0 \sim 0.3$ K, where non-Fermi-liquid (NFL) behavior in the specific heat and electrical resistivity is observed. This logarithmic temperature variation manifested in the $\Gamma_3$-symmetry quadrupolar susceptibility is consistent with the theoretical prediction of the quadrupolar Kondo effect by D. L. Cox [1]. On the other hand, the pure compound YIr$_2$Zn$_{20}$ without 4$f$-electron contributions shows nearly no change in its elastic constants evidencing negligible phonon contributions. In addition, clear acoustic de Haas-van Alphen (dHvA) oscillations in the elastic constant were detected for both compounds on applying magnetic field. This is mainly interpreted as contribution from the Fermi surface of YIr$_2$Zn$_{20}$.

**KEYWORDS:** quadrupolar Kondo effect, non-Kramers doublet, acoustic dHvA effect, ultrasound, quadrupolar susceptibility, elastic constant

## 1. Introduction

Concomitant non-Fermi-liquid (NFL) behavior of physical quantities due to an unconventional Kondo effect has been discussed [2]. Some compounds that evidence NFL behavior have been recognized as exhibiting the quadrupolar Kondo effect (QKE) [1], which is understood to be the two-channel version of the multi-channel Kondo effect [3]. The diluted Pr system of the cubic Pr-based non-Kramers doublet system, Y$_{1-x}$Pr$_x$Ir$_2$Zn$_{20}$, has recently been studied systematically. Pr-diluted systems display possible single-site quadrupolar Kondo behavior, which appears as NFL behavior in specific heat and resistivity [4]. On the other hand, the non-diluted compound PrIr$_2$Zn$_{20}$ ($x$ = 1) shows a localized quadrupolar order of the $\Gamma_3$ non-Kramers doublet ground-state [5] and superconductivity [6]. In this paper, we show acoustic signatures of the single-site quadrupolar Kondo effect in the Pr dilute-limit Y$_{0.966}$Pr$_{0.034}$Ir$_2$Zn$_{20}$. As the main result the logarithmic elastic response for the quadrupolar Kondo effect has been reported in a previous paper [7], that we outline in section 3.1. We then focus on the analysis of background subtraction (section 3.2), acoustic de Haas-van Alphen (dHvA) effect (section 3.3), and comparison of the fit parameters for crystalline electric field (CEF) analysis of the quadrupolar susceptibility with the mother compound $PrIr_2Zn_{20}$ (section 3.4) to corroborate the reliability of our interpretation of the logarithmic temperature dependence.

## 2. Experimental Details

Single-crystalline $Y_{1-x}Pr_xIr_2Zn_{20}$ samples were grown using the Zn-self-flux method as described in previous papers [4, 8]. The sample quality was checked using the powder x-ray diffraction (XRD) technique testing another sample from the same batch. The results showed no detectable vacancies on the Y or Pr sites because all the Rietveld analyses of the power XRD pattern converged when the filling factor of Y and Pr was set to 100%. For the present sample, the nominal Pr composition of $x = 0.034$ was confirmed by considering the magnetization data at 1.8 K and the electron-probe microanalysis (EPMA) results. We also checked the spatial distribution of the Pr composition using scanning electron microscopy (SEM) and conclude that the Pr composition in each piece of the sample is homogeneous, i.e., the inhomogeneity is within ~1%. Sample dimensions of the rectangular parallelepiped are $2.954 \times 2.771 \times 2.440$ mm$^3$ for the $x = 0.034$ sample and $3.538 \times 2.106 \times 1.428$ mm$^3$ for the $x = 0$ ($YIr_2Zn_{20}$) sample for [110]-[1$\bar{1}$0]-[001], respectively. A pair of 100 μm thick $LiNbO_3$ wafers were fixed on the sample surfaces with Room Temperature Vulcanizing (RTV) silicone and used as ultrasonic transducers. The ultrasonic velocity and attenuation coefficient were observed using the phase-comparative method. The low-temperature measurements were performed using a $^3$He-$^4$He top-loading dilution refrigerator.

## 3. Results

### 3.1 Acoustic Signature of Quadrupolar Kondo Effect: Logarithmic Temperature Dependence of $(C_{11}-C_{12})/2$

Figure 1 displays the elastic constant $(C_{11}-C_{12})/2$ (= $C_v$) of $Y_{0.966}Pr_{0.034}Ir_2Zn_{20}$ as a function of temperature at zero magnetic field and 14 T applied along the [001] axis. Additionally, the temperature dependence of $(C_{11}-C_{12})/2$ of $YIr_2Zn_{20}$ under the same conditions is displayed in Fig. 1 for comparison as a reference to the phonon background. The $(C_{11}-C_{12})/2$ mode of $Y_{0.966}Pr_{0.034}Ir_2Zn_{20}$, which corresponds to the $\Gamma_3$-symmetry electric-quadrupolar response, shows a logarithmic temperature dependence in the low-magnetic-field region below ~0.3 K ($\Delta C_v/C_v$ ~0.1%). We compared the relative change of the low-temperature softening $\Delta C_v/C_v$ below 1 K between the pure compound ($x = 0$) and the dilute system ($x = 0.034$), with its magnetic field variation within 0–14 T. The change in the pure compound ($\Delta C_v/C_v$ ~0.002%), caused by the phonon contribution in $YIr_2Zn_{20}$, is ~10 times smaller than the softening in the diluted $x = 0.034$ sample at 14 T ($\Delta C_v/C_v \sim 0.017\%$). Thus, the phonon contribution from $YIr_2Zn_{20}$ is negligible. On the other hand, the Curie-type divergence of the elastic constant for $x = 0.034$ in the high-temperature region (above ~1 K) indicates that the Pr ions in this diluted system have a non-Kramers ground-state doublet. The dotted curves in the main panel of Fig. 1 show the calculated elastic constant based on the local quadrupolar susceptibility of the localized 4$f$-electron system applying a crystalline electric field (CEF) model (explained in section 3.4) [7]. The calculation deviates from the experimental data at 0 T; while it cannot reproduce the logarithmic temperature dependence at low temperatures, it can do so sufficiently above 3 T [7]. The inset of Fig. 1 shows the background-subtracted data for 0, 1, and 2 T. Details of the additional background subtraction, which is due to the dHvA effect and additional phonon contributions from Pr ions, are explained in the next section. This logarithmic temperature variation, manifested in the $\Gamma_3$-symmetry quadrupolar susceptibility, is consistent with the theoretical prediction of the QKE by D. L. Cox [1]. These observations evidence the single-site QKE, as previously suggested based on specific-heat data. In the inset of Fig. 1, the results of regression analyses with the temperature variation proportional to $+\ln T$ and $-\sqrt{T}$ are shown as a black solid line and a blue dashed curve, respectively. The $-\sqrt{T}$-law of several physical quantities, such as specific heat and multipolar susceptibility, was predicted based on another theory, by A. Tsuruta and K. Miyake [9], regarding a two-channel Anderson lattice model with a strong correlation effect. This law reproduced the NFL behavior of the electrical resistivity in the present sample. However, the present ultrasonic results of $(C_{11}-C_{12})/2$ do not follow the $-\sqrt{T}$ law, rather, they follow $+\ln T$. This contradiction still remains and is open to questions.

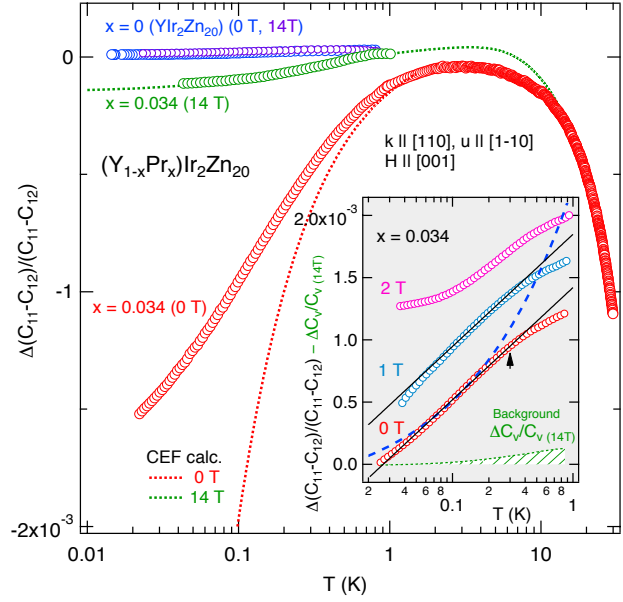

**Fig. 1.** Relative change of the elastic constant $(C_{11}-C_{12})/2$ of $Y_{1-x}Pr_xIr_2Zn_{20}$ as a function of temperature at zero magnetic field and 14 T for $H \parallel [001]$. Dotted curves represent the calculated elastic constant with the CEF model (see text). The inset shows the temperature dependence of the background-subtracted elastic constant below 1 K for $H = 0$, 1, and 2 T. The black solid line and blue dashed curve indicate the fit of $+\ln T$ and $\sqrt{T}$ dependences, respectively.

*3.2 Background Subtraction of the de Haas-van Alphen Effect and Possible Off-center Tunneling*

Background subtraction of the elastic response is important for analysis of the logarithmic temperature dependence of the elastic constant in the low-magnetic-field region, as shown in the inset of Fig. 1. Here, we provide additional information regarding the background estimation to clarify the data-subtraction process. A schematic illustration of the background subtractions is shown in Fig. 2. The value of the constant background $C_{\Gamma 3}^{\text{dHvA}}$ is defined by the dHvA oscillation amplitude, which is estimated from the magnetic-field dependence of the elastic constant at the lowest temperature, as shown in Fig. 3. For example, the constant background $C_{\Gamma 3}^{\text{dHvA}} = +0.005$ J/m³ for 14 T, is defined based on the dHvA oscillation amplitude at 14 T and the lowest temperature 40 mK.

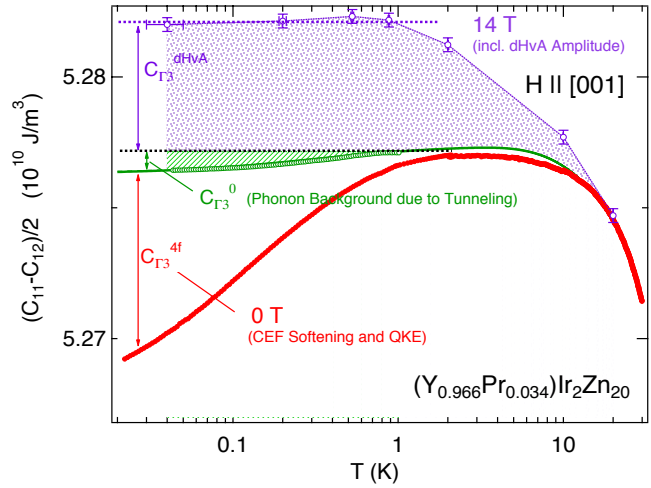

**Fig. 2.** Schematic illustration of the background subtraction for the elastic constant $(C_{11}-C_{12})/2$ of Y$_{0.966}$Pr$_{0.034}$Ir$_2$Zn$_{20}$. Open circles with error bars represent the temperature dependence of the elastic constant, which was converted from the magnetic-field dependence at fixed temperatures as shown in Fig. 3. The blue and green hatched areas represent the estimated dHvA contributions and phonon contributions (including quantum tunneling) $C_{\Gamma 3}^{0}$, respectively, as the backgrounds for the CEF analysis.

The dHvA oscillation amplitude was $\Delta C_{\text{v}}/C_{\text{v}} \sim 1.4\text{e-}4$ at $\sim 14$ T below $\sim 1$ K. This value is much smaller and negligible compared to the total amplitude $\Delta C_{\text{v}}/C_{\text{v}} \sim 1.2\text{e-}3$ of the logarithmic softening in the temperature dependence. Figure 3 displays the dHvA signal and its Fourier spectrum of YIr$_2$Zn$_{20}$ ($x = 0$) up to 17 T under several fixed temperatures. The change in the dHvA oscillation amplitude of the pure compound below 960 mK, $\Delta C_{\text{v}}/C_{\text{v}} = 2.0\text{e-}5$ at 3 T, is two orders of magnitude less than the change in the $+\ln T$ dependence of the dilute system at 0 T in the same temperature range. Therefore, the temperature dependence of the dHvA contribution below 1 K can be considered negligible especially in the low-magnetic-field region, and the constant background subtraction of the dHvA effect in the current case does not affect the results presented in Fig. 1. In the main panel of Fig. 1, we intentionally subtracted the constant background, which is originating in the dHvA effect, from the temperature dependence of $(C_{11}-C_{12})/2$ under magnetic field for easy comparison of the CEF and QKE contributions.

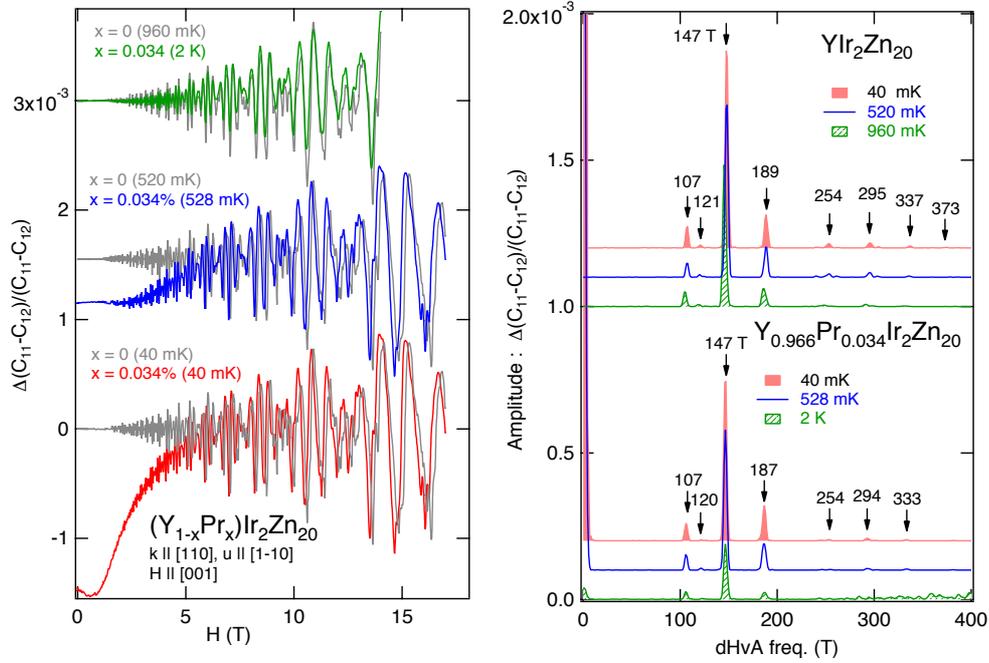

**Fig. 3.** (Left) Magnetic-field dependence of the elastic constant $(C_{11}-C_{12})/2$ of $Y_{0.966}Pr_{0.034}Ir_2Zn_{20}$ and $YIr_2Zn_{20}$. (Right) Fourier spectrum of the acoustic dHvA signals.

In the inset of Fig. 1, the data measured at 14 T as the magnetically robust phonon contribution are additionally subtracted. Intriguingly, this phonon contribution is more enhanced in the Pr-diluted system compared to the pure compound with $x = 0$; this is probably due to a lattice instability originating from the local oscillations of the Pr or Zn ions, known as 'quantum tunneling' [10, 11]. The quantum tunneling (or off-center tunneling) is defined as a local Einstein-phonon-like, quantum oscillation of the atom through the potential hills between potential minima around the high-symmetry sites. This phenomenon has generally been found in glasses (as two-level system) [12] and some cage-structured compounds, such as clathrate metals [10], filled-skutterudite compounds [13], and the present $RT_2Zn_{20}$ (R = Y, La or Pr, T = Rh or Ir) compounds. The symmetrical, off-center mode couples to the appropriate, symmetrized strain (deformation potential) induced by ultrasound, and causes a magnetically-insensitive Curie-type softening of the elastic constant at low temperatures. Such quantum-tunneling contribution has not been discussed in the present 1-2-20 systems, and thus, further investigation will be necessary to acquire knowledge on this issue. From the above background subtraction, we obtain a more pronounced $+\ln T$ dependence below $T_0$ toward the lowest temperatures.

*3.3 Acoustic de Haas-van Alphen Effect*

The observed clear acoustic dHvA signals prove the high quality of the investigated single crystals. We analyzed these dHvA signals in the conventional manner. The left panel of Fig. 3 shows the magnetic field dependence of the elastic constant $(C_{11}-C_{12})/2$ of $Y_{0.966}Pr_{0.034}Ir_2Zn_{20}$ and $YIr_2Zn_{20}$. Both signals are similar, except for additional CEF and QKE contributions at lower temperatures and magnetic-field regions for the diluted system. Thus, it is reasonable to assume that the acoustic dHvA signal in the dilute system mainly originates from the signal of the pure compound $YIr_2Zn_{20}$.

Fourier spectra of the acoustic dHvA signals are shown in the right panel of Fig. 3. There is little variation among these spectra. Acoustic dHvA oscillations with frequencies $F$ of 147, 187, 107, 294, 254, 333, and 120 T (decreasing order of oscillation amplitude) were clearly observed as peaks of the spectra for both the pure compound and the diluted system. These values are comparable to the dHvA frequencies of $PrIr_2Zn_{20}$ ($F$ = 150 T, $m_c^*$ = 0.37$m_0$ for band 269) and $LuIr_2Zn_{20}$ ($F$ = 125 T, $m_c^*$ = 0.25$m_0$ for band 281) previously reported, which were measured using the standard field-modulation method (magnetization) [14]. However, some of the obtained dHvA frequencies for the dilute system and the pure compound differ slightly ($\pm$ 2 T). The dHvA frequencies of the diluted system are also slightly different from those reported in the previous paper [7], because the Fourier transformation was performed for different magnetic field ranges between 4.5 and 17 T in the present analysis, while the previous Fourier transformation was performed between 2 and 14 T. Here, the average deviations of $\Delta F \sim \pm$ 2 T among these data can be considered as the accuracy limit of the present analysis. We can approximately estimate the cyclotron masses from the temperature dependence of the oscillation amplitude. For instance, we obtain $m_c^*$ = 0.39$\pm$0.03$m_0$ for the 147 T, $m_c^*$ = 0.4$\pm$0.1$m_0$ for the 189 T, and $m_c^*$ = 0.6$\pm$0.3$m_0$ for the 107 T orbit of the pure compound $YIr_2Zn_{20}$. These values are of the same order as those estimated experimentally and theoretically for $PrIr_2Zn_{20}$ and $LuIr_2Zn_{20}$ [14].

Further investigations should include angular-dependent dHvA measurements and comparison of the results to band-structure calculations.

*3.4 Comparison of the CEF Parameters between the Mother Compound (x =1) and the Dilute System*

In this section, we demonstrate the process of exploring the CEF parameters for the diluted system in order to show the fact that the amount of deviation from the low-temperature softening may not necessarily scale with the Pr composition of the dilute system when the inter-site interaction drastically goes to zero for $x \rightarrow 0$. The panels in Fig. 4 illustrates the relative change in the elastic constant $(C_{11}-C_{12})/2$ ($= C_v$), which is normalized at 1 K, of $(Y_{1-x}Pr_x)Ir_2Zn_{20}$, $x$ = 1 and 0.034 with logarithmic temperature scale (left) and linear scale (right). When we compare the magnitude of the softening down to $\sim T_0$ for the $x$ = 1 and $x$ = 0.034 sample, the change $\Delta C_v/C_v$ does not perfectly scale with the Pr concentration. However, the lack of the antiferroquadrupolar interaction in the diluted system causes an effective enhancement of the quadrupolar response.

The $\Gamma_3$-symmetry quadrupolar susceptibility is also shown in Figs. 4, which is calculated based on the single-ion quadrupolar susceptibility using the quadrupolar operator $O_2^2 = J_x^2 - J_y^2$ and the previously reported CEF parameters for PrIr$_2$Zn$_{20}$, $W = -1.219$ K and $X = 0.537$, for the non-Kramers ground-state doublet with the $\Gamma_3$ symmetry and the first excited triplet with the $\Gamma_4$ symmetry (at ~25.3 K in zero magnetic field) [7]. They are represented by the green and red solid curves in Figs. 4.

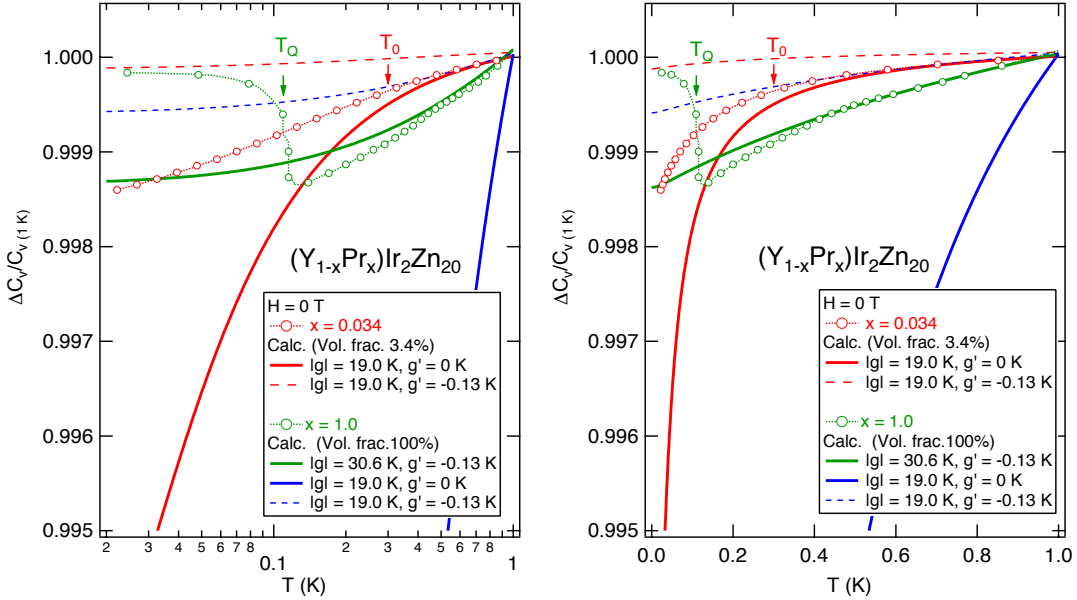

**Fig. 4.** Comparison of experimental results and calculations of the elastic constant $C_v = (C_{11}-C_{12})/2$, shown as the relative change normalized by the value at 1 K, with logarithmic (left) and linear (right) temperature axes. PrIr$_2$Zn$_{20}$ ($x = 1$) data [5] are also displayed for comparison.

The present CEF analysis of the diluted system is done using several values for the coupling constant $|g|$ and $g'$. A finite (negative) value of the inter-site quadrupole-quadrupole interaction, $g' = -0.13$ K, for the mother compound PrIr$_2$Zn$_{20}$ ($x = 1$) has been reported by Ishii *et al.* [5]. On the other hand, for the non-magnetic Y-diluted samples, the coupling constant of the inter-site quadrupolar-quadrupolar interaction should be set to $g' = 0$, since there is negligibly small inter-site quadrupolar interaction between the well-separated Pr ions in the system.

As a test, we fixed the quadrupolar-strain interaction to $|g| = 19$ K with inter-site interaction $g' = -0.13$ K (dashed red and blue curves), and $g' = 0$ K (solid red and blue curves). The calculated results are scaled to the volume fraction of the Pr ion. In searching for the optimized CEF parameters in the present analysis, we fixed the volume and then changed the coupling constants, $|g|$ and $g'$. Generally, the localized-electron model should be excluded well below the attempt temperature $T_K$ (~$T_0$ in the present case) of the single-channel and/or multichannel Kondo effect, since the local multipolar moment will be screened out by conduction electrons. We searched for the best fit to reproduce the temperature and magnetic field dependence of the elastic constant above $T_0$. Thus, we give due consideration due to the CEF analysis in the high-magnetic-field regions as it is plausible that the local 4$f$-electron model cannot be applied in the low-temperature region (below $T_0$ ~0.3 K), where the QKE is dominant. Finally, we obtained $|g| = 19$ K, $g' = 0$ K as the best fit. The above comparisons indicate that the magnitude of the low-temperature softening depends on the trading-off relationship between the volume fraction of Pr and antiferro-quadrupolar interaction $g'$.

## 4. Conclusion

We reported the elastic constant $(C_{11}-C_{12})/2$ of the Pr-diluted system $Y_{0.966}Pr_{0.034}Ir_2Zn_{20}$, which is a candidate compound in which the single-site quadrupolar Kondo effect (QKE) is realized. The data were compared to the reference non-4$f$-electron system, $YIr_2Zn_{20}$, revealing the background contribution of the elastic moduli. From the present results and analysis, we conclude: i) The non-Kramers ground-state doublet with $\Gamma_3$ symmetry in the dilute system, which is crucial for the QKE, is proven by the present ultrasound results. ii) The channel number of the present Kondo effect must be two, as directly evidenced by our ultrasonic result, which clearly shows a logarithmic temperature dependence of the $\Gamma_3$ quadrupolar susceptibility due to the "single-site" multi-channel Kondo effect. Further measurements of samples with different Pr concentrations are needed to confirm a possible QKE scaling effect and to determine systematic changes in the CEF parameters and phonon contributions.

## Acknowledgment

This work was supported by the HLD at HZDR, member of the European Magnetic Field Laboratory (EMFL); JSPS KAKENHI Grants Nos. JP15KK0169, JP18H04297, JP18H01182, JP17K05525, JP18KK0078, JP15KK0146, JP15H05882, JP15H05885, JP15H05886, JP15K21732, JP19J10235, and the Strategic Young Researcher Overseas Visits Program for Accelerating Brain Circulation. Y. Y. was supported by JSPS Research Fellowships for Young Scientists. T. Y. would like to thank J. Klotz for supporting the operation of the dilution refrigerator at HZDR.